\documentclass[%
 aps,
 pre,%
 amsmath,amssymb,
reprint,%
groupedaddress,
superscriptaddress,
author-numerical,%
]{revtex4-2}

\usepackage{graphicx}
\usepackage{dcolumn}
\usepackage{bm}
\usepackage[mathlines]{lineno}
\usepackage{amsmath}
\usepackage{url}

\usepackage{booktabs}
\usepackage{rotating}
\usepackage{multirow}
\usepackage{subfigure}
\usepackage[normalem]{ulem}
\usepackage{amsmath}
\usepackage[usenames,dvipsnames,svgnames,table]{xcolor}
\usepackage{booktabs}
\usepackage{rotating}
\usepackage{multirow}
\usepackage{subfigure}
\usepackage{tabularx}

\usepackage{color}
\usepackage{hyperref}
\hypersetup{
     colorlinks   = true,
     citecolor    = blue,
     linkcolor    = blue,
     urlcolor     = blue
}
\usepackage{diagbox}

\newcommand{\AK}{\textcolor{ForestGreen}}

\begin{document}

\title[Sample title]{General scaling laws of space charge effects in field emission}

\author{A. Kyritsakis}
\email{akyritsos1@gmail.com}
\affiliation{Helsinki Institute of Physics and Deparment of Physics, University of Helsinki, PO Box 43 (Pietari Kalmin katu 2), 00014 Helsinki, Finland}
\affiliation{Institute of Technology, University of Tartu, Nooruse 1, 50411 Tartu, Estonia}

\author{M. Veske}%
\affiliation{Helsinki Institute of Physics and Deparment of Physics, University of Helsinki, PO Box 43 (Pietari Kalmin katu 2), 00014 Helsinki, Finland}


\author{F. Djurabekova}
\affiliation{Helsinki Institute of Physics and Deparment of Physics, University of Helsinki, PO Box 43 (Pietari Kalmin katu 2), 00014 Helsinki, Finland}
\affiliation{National Research Nuclear University MEPhI, Kashirskoye sh. 31, 115409 Moscow, Russia}

\date{\today}

\begin{abstract}
The characteristics of field electron and ion emission change when the space charge formed by the emitted charge is sufficient to suppress the extracting electric field.
This phenomenon is well described for planar emitting diodes by the one dimensional (1D) theory.
Here we generalize for any 3D geometry by deriving the scaling laws describing the field suppression in the weak space charge regime.
We propose a novel corrected equivalent planar diode model, which describes the space charge effects for any geometry in terms of the 1D theory, utilizing a correction factor that adjusts the diode's scaling characteristics.
We then develop a computational method, based on the Particle-In-Cell technique, which solves numerically the space charge problem. We validate our theory by comparing it to both our numerical calculations and existing experimental data, either of which can be used to obtain the geometrical correction factor of the corrected equivalent planar diode model.

\end{abstract}

\keywords{Electron emission, Field emission, space charge, Particle In Cell}
\maketitle

\section{Introduction}

The current extracted from an electron emitting cathode or an ion emitting anode can be increased by applying higher electric fields, increasing the emitter temperature or irradiating it with light. However, the current density cannot increase beyond a certain limit, due to the space charge effect \cite{Child_SC,langmuir1913effect}, i.e. the suppression of the extracting field due to the space charge formed by the emitted particles.
The space charge (SC) effect plays a very significant role in all forms of electron and ion sources \cite{bormann2010tip, dyke1956field, chen2009space,fursey2007field, mair1982space, BarbourSC,radlicka2008coulomb} and is of paramount importance for the understanding of the ignition of vacuum arcs \cite{Dyke1953Arc, wagenaars2007measurements, kyritsakis2018thermal,veske2019dynamic}.

Despite this importance, there is no general three dimensional (3D) theory describing the SC effect for curved emitter geometries, mainly due to the high complexity of the problem.
In order to calculate the SC effects one has to solve self-consistently three coupled problems: the Poisson equation, the continuity equation, and the electron emission equation.
This is possible by utilizing various numerical methods, such as Particle-In-Cell (PIC) \cite{feng2006transition,Uimanov2011, zhu2015space}, molecular dynamics \cite{torfason2016molecular}, or the point charge method \cite{jensen2010space}.

Analytical solutions of the SC problem exist only for 1D geometries, where the continuity equation has a simple solution.
For instance, Child \cite{Child_SC} and Langmuir \cite{langmuir1913effect} solved the problem for a planar diode geometry and the special case of fully SC-limited charge flow, i.e. with a boundary condition of zero electric field at the emitting electrode. 
To distinguish this case from the general problem with a non-zero field at the emitter, we shall call it special SC problem. 
Later on, Langmuir and Blodgett developed analytical solutions of the special SC problem in the form of a series for non-planar --but, still 1D-- geometries of concentric cylindrical \cite{langmuir1923currents} and spherical \cite{langmuir1924currents} electrodes. 
In a recent work \cite{zhu2013novel}, it was shown that both problems yield general scaling laws similar to the planar diode.

The general SC problem, which is relevant for field electron and ion emission, was solved for the planar case by Stern {\it et. al.} \cite{stern1929further}.
This solution has been widely used as a reference model to estimate the SC effects on field emission \cite{goncalves2001experimental,rokhlenko2010space,ForbesSpace,chen2009space}, due to its simplicity and intuitiveness.
The connection of the planar diode model to an experimental emitter geometry is typically done via the equivalency introduced by Barbour \textit{et. al.} \cite{BarbourSC}, i.e. by assigning the real values of voltage and cathode field to a planar diode.

However, PIC simulations \cite{Uimanov2011,zhu2015space,veske2019dynamic} have clearly shown that this equivalence is not always valid, as it tends to significantly overestimate the SC suppression of the field, and thus may lead to wrong values of emission current from a 3D emitter. 
In this work, we tackle this problem by developing a general 3D theory for space charge suppressed emission.
In section \ref{sec:general} we derive the general scaling laws for the emission behavior in the weak SC regime, by introducing the corrected equivalent planar diode (CEPD) model and showing that any 3D emitter is equivalent, regarding SC, to a planar diode of certain characteristics, determined by a single geometry-dependent correction factor $\omega$.
In section \ref{sec:sphercyl} we derive $\omega$ for the spherical and cylindrical diode problems.
In section \ref{sec:computational} we describe a general numerical method to obtain \AK{$\omega$} for any 3D electrode geometry.
We finally validate our theory by comparing to both numerical calculations and existing experimental data in section \ref{sec:results}, showing that the correction factor can be obtained for a certain geometry by fitting to either numerical calculations or experimental data.

\section{Theory}
\label{sec:theory}
\subsection{General formulation}
\label{sec:general}
The standard formulation of the space charge problem adopted already since Child's work \cite{Child_SC}, assumes a continuous charge density distribution of the emitted charge $\rho(\mathbf{r})$ and zero initial velocity for the emitted charged particles.
Some studies have considered the case of non-zero initial kinetic energy \cite{jensen2010space}, but since the latter is of the order of a few eV, we shall consider it negligible for the cases of emission under high electric field.
Under these assumptions, the Poisson equation becomes
\begin{equation} \label{eq:Poisson}
\nabla^2 \Phi =kJ(\mathbf{r})\Phi^{-1/2} \textrm{,}
\end{equation}
with boundary conditions $\Phi = 0$ at the emitter and $\Phi = V$ at the collector electrodes.
Here $\Phi$ is the electrostatic potential, $k=\epsilon_0^{-1}\sqrt{m/2q}$ is a constant that depends on the mass to charge ratio $m/q$ of the emitted particles and the dielectric constant $\epsilon_0$; $V$ is the applied voltage.
The current density distribution $J(\mathbf{r})$ obeys the continuity equation $\nabla \cdot \mathbf{J} = 0$, with boundary condition $\mathbf{J}(\mathbf{r}_s) = J_s(\mathbf{r}_s)\hat{n}(\mathbf{r}_s)$, where $J_s(\mathbf{r}_s)$ is the emitted current density and $\hat{n}(\mathbf{r}_s)$ is the normal unit vector at the emitting surface point $\mathbf{r}_s$. 
In order to obtain $\mathbf{J}$ and $\Phi$, the above equations have to be solved self-consistently, along with the surface emission laws that give $J_s(\mathbf{r}_s)$ as a function of the local electric field $F(\mathbf{r}_s)$.

The above problem cannot be solved analytically, apart from the planar case, where $J$ is constant. 
In this case, the solution of the 1D Poisson equation for a planar emitting diode with a gap distance $d$, voltage $V$, current density $J$ and cathode field $F$, yields \cite{stern1929further}
\begin{equation} \label{eq:stern}
    6(kJ)^2d-F^3 = \left(2kJ\sqrt{V}-F^2 \right) \sqrt{4kJ\sqrt{V}+F^2} \textrm{.}
\end{equation}
In case the current density $J$ supplied at the cathode depends on the cathode field $F$ (e.g. field electron or ion emission), eq. \eqref{eq:stern} needs to be solved self-consistently with the law that gives $J(F)$, e.g. the Fowler-Nordheim (FN) law \cite{FN1928,MurphyG} or its modern generalizations \cite{Jensen2006,KXnonfn, KXGTF,GETELECpaper}.
Barbour \textit{et. al.} \cite{BarbourSC} used approximate methods to achieve this self-consistency, while modern iterative numerical methods allow to find a more accurate solution at relatively low computational costs \cite{kyritsakis2018thermal}.

Eq.~\eqref{eq:stern} can be simplified by introducing the reduced dimensionless variables \cite{goncalves2001experimental,ForbesSpace,rokhlenko2010space} $\theta \equiv Fd/V=F/F_L$ and $\zeta \equiv kJd^2V^{-3/2} = kJd^{1/2}F_L^{-3/2} = kJV^{1/2} / F_L^2$, yielding
\begin{equation} \label{eq:Goncalves}
    3\theta^2(1- \theta) = \zeta(4 - 9\zeta) \textrm{.}
\end{equation}
In \eqref{eq:Goncalves}, $\theta$ is the ``field reduction factor'', i.e. the factor by which the field has been reduced from the ``Laplace field'' $F_L \equiv V/d$ due to the SC.
$\zeta$ is indicative of the ``space charge strength'', since it is evident from \eqref{eq:Goncalves} that $\theta$ reduces from 1 to 0 ($F$ from $F_L$ to 0) as $\zeta$ increases from 0 to $4/9$, where the Child law \cite{Child_SC} limit occurs.
Eq.~\eqref{eq:Goncalves} can be solved analytically \cite{rokhlenko2010space}; yet, it is more convenient to express the physical solution in a perturbation series around $\zeta=0$, yielding
\begin{equation}
\label{eq:jseries}
    \theta = 1- \frac{4}{3}\zeta - \frac{5}{9}\zeta^2 - \frac{16}{27}\zeta^4 + \cdots \textrm{.}
\end{equation}  
Truncating this series at the linear term yields a good approximation with an error of less than 10\% for $\zeta < 0.25$ and $\theta > 0.6$.
This ``weak SC regime'' covers most practical cases in field electron and ion emission, where typically  $J_s \lesssim 10^{12} \textrm{A/m}^2$) \cite{dyke1956field,fursey2007field}.
The generalization of this scaling behavior of $\theta(\zeta)$ for non-planar geometries is the main purpose of this article.

Consider an emitter with a general 3D geometry and a point of interest $\mathbf{r}_s$ at the emission surface, with local field $F$.
We shall express the Poisson equation in terms of the reduced variables $\phi \equiv \Phi / V$, $J(\mathbf{r})=J_s \xi(\mathbf{r})$, where $J_s = J(\mathbf{r}_s)$ and $\xi(\mathbf{r})$ is a unitless variable expressing the variation of the current density from $\mathbf{r}_s$.
Assuming that the distribution of the surface emission $J_s$ does not vary significantly and using the linearity of the continuity equation, we can approximate that $\xi(\mathbf{r})$ depends only on the emitter geometry and not on $J_s$.
Finally, we use the reduced space coordinate $\tilde{\mathbf{r}} = \mathbf{r} / \chi$, where $\chi$ is the ``conversion length'' $\chi \equiv V / F_L$ ($F_L$ is the Laplace ($J=0$) field at $\mathbf{r}_s$).
Equation~\eqref{eq:Poisson} becomes
\begin{equation} \label{eq:pscaled}
    \tilde{\nabla}^2 \phi(\tilde{\mathbf{r}}) = \left( \frac{kJ_s\chi^2}{V^{3/2}} \right) \frac{\xi(\tilde{\mathbf{r}})}{\sqrt{\phi(\tilde{\mathbf{r}})}} \textrm{,}
\end{equation}
where $\tilde{\nabla}$ denotes derivatives with respect to $\tilde{\mathbf{r}}$. 
Note that the reduced position $\tilde{\textbf{r}}$ is scale invariant, i.e. it remains unchanged under a geometrical scaling $\textbf{r} \rightarrow a \textbf{r}$, for any scaling factor $a$, as $F_L$ scales as $1/a$ and $\chi$ as $a$.

The significance of the above equation becomes evident in view of the correspondence of the parameter in the parenthesis $\zeta \equiv kJ_s \chi^2V^{-3/2}=kJ_s V^{1/2}/F_L^{2}$ with the space charge strength of the planar diode.
The classical planar diode equivalence~\cite{BarbourSC,ForbesSpace} corresponds to inserting the above $\zeta$ to eq.~\eqref{eq:jseries}, or equivalently substituting $d$ with $\chi$ in eq.~\eqref{eq:stern}.
Now we shall derive an asymptotic expansion similar to eq.~\eqref{eq:jseries} for a general 3D geometry, showing that the above equivalence is not valid, and propose a new one that holds up to the first order on $\zeta \ll 1$.

We express eq.~\eqref{eq:pscaled} in its integral form, utilizing the Green's function for Dirichlet problems \cite{melkinov2012green} $G(\tilde{\mathbf{r}}, \tilde{\mathbf{r}}')$, i.e. the solution of the boundary value problem (BVP) $\tilde{\nabla}_{\mathbf{r}}^2 G(\tilde{\mathbf{r}}, \tilde{\mathbf{r}}') = \delta(\tilde{\mathbf{r}}- \tilde{\mathbf{r}}') \textrm{ on } \Omega \textrm{, } G(\tilde{\mathbf{r}}, \tilde{\mathbf{r}}') = 0 \textrm{ on } \partial\Omega$, with $\delta(\cdot)$ being Dirac's functional and $\Omega$ the vacuum domain.
Then the solution can be written in the form of a Fredholm integral equation as
\begin{equation} \label{eq:pinteg}
    \phi(\tilde{\mathbf{r}}) = \phi_0(\tilde{\mathbf{r}}) + \zeta \int_{\Omega} \frac{G(\tilde{\mathbf{r}}, \tilde{\mathbf{r}}') \xi(\tilde{\mathbf{r}}')}{\sqrt{\phi(\tilde{\mathbf{r}}')}} d^3\tilde{\mathbf{r}}' \textrm{,}
\end{equation}
where $\phi_0(\tilde{\mathbf{r}})$ is the Laplace solution ($\zeta=0$).
For $\zeta \ll 1$, $\phi(\tilde{\mathbf{r}})$ can be expanded in an asymptotic series using the Adomian decomposition method \cite{wazwaz2011linear}, as
\begin{equation} \label{eq:adomian}
    \phi(\tilde{\mathbf{r}}) = \phi_0(\tilde{\mathbf{r}}) + \zeta \int_{\Omega} \frac{G(\tilde{\mathbf{r}}, \tilde{\mathbf{r}}') \xi(\tilde{\mathbf{r}}')}{\sqrt{\phi_0(\tilde{\mathbf{r}}')}} d^3\tilde{\mathbf{r}}' + O(\zeta^2) \textrm{.}
\end{equation}
By taking the gradient we obtain a similar expansion for the field reduction factor $\theta = F / F_L$
\begin{align}       \label{eq:omega}
\begin{split}
    \theta &= 1 - \frac{4}{3} \omega \zeta + O(\zeta^2) \textrm{, with} \\
    \omega & \equiv \frac{3}4 \int_{\Omega}  \left.\tilde{\nabla}_{\mathbf{r}} G(\tilde{\mathbf{r}}, \tilde{\mathbf{r}}')\right\vert_{\tilde{\mathbf{r}}_s} \frac{\xi(\tilde{\mathbf{r}}')}{\sqrt{\phi_0(\tilde{\mathbf{r}}')}} d^3\tilde{\mathbf{r}}' \textrm{.}
\end{split}
\end{align}

The above equation demonstrates the inadequacy of the standard planar diode equivalence of Refs~\cite{BarbourSC,ForbesSpace}.
The scaling of $\theta(\zeta)$ for $\zeta \ll 1$ is the same as the planar diode only if $\omega = 1$,
which does not hold in general.
However, $\omega$ can be incorporated as a correction to the planar model.
Thus, for a given geometry and a surface point $\mathbf{r}_s$, we define the corrected equivalent planar diode model, for which the SC strength is $\zeta_{c} = \omega kJ_s\chi^2V^{-3/2}=\omega kJ_s\sqrt{\chi}/F_L^{3/2}$.
Practically, the CEPD model can be used by replacing $\zeta = \zeta_c$ in eq.~\eqref{eq:Goncalves}, or using the CEPD gap distance $d_c = \omega^2 \chi$ and voltage $V_c = \omega^2V$ in eq.~\eqref{eq:stern} and solving for $F$.
Alternatively, one can substitute directly $V$ and $d=\chi$ in ~\eqref{eq:stern}, as in Refs \cite{BarbourSC, ForbesSpace}, but use a corrected current density value $J_c = \omega J_s$.
An important property of the geometry correction factor $\omega$ is its scale invariance, resulting from the fact that all variables in eq. \eqref{eq:omega} are invariant to geometrical scaling. 
In other words, $\omega$ does not depend on the absolute size of the electrodes, but only on their relative shape, thus simplifying its calculation and tabulation for various geometries.

The usage of a correction factor such as $\omega$ has been proposed by Forbes~\cite{ForbesSpace}.
Here we define and calculate it rigorously.
$\omega$ can be formally calculated for a given emitter geometry and point $\mathbf{r}_s$ using \eqref{eq:omega}, which is though quite cumbersome.
It is practically more convenient to calculate $\theta(\zeta)$ numerically (e.g. by PIC) for a given geometry, and fit $\omega$ to the results, thus yielding a more general and computationally cheap approximate solution of the SC problem.
Before advancing to such simulations, we shall focus on two diode geometries, for which the CEPD can be obtained analytically, giving useful physical insights.

\subsection{Analytical solutions for spherical and cylindrical geometries}
\label{sec:sphercyl}

The geometries of the spherical and cylindrical diodes, for which the electrodes are two concentric spheres or cylinders correspondingly, have attracted theoretical interest since the early SC studies \cite{langmuir1923currents,langmuir1924currents,poplavskii1950potential, aizenberg1954role}. 
Langmuir and Blodgett \cite{langmuir1923currents,langmuir1924currents} solved the specific SC problem for both geometries.
Furthermore, for the spherical case, which has been recently used to model sharp field emitters \cite{chen2009space},  Aizenberg \cite{aizenberg1954role} obtained an approximate solution for the general SC problem.
Here we apply the CEPD model and calculate $\omega$ for both the spherical and cylindrical geometries, thus obtaining the corresponding scaling laws for the weak SC regime.

The resulting expressions for the CEPD correction factor $\omega$ are
\begin{align}       \label{eq:omega_sph_cyl}
\begin{split}
    \omega^{(S)} & = \frac{3}{4} \frac{ \left(2- \frac{1}{\tilde{r}} \right) \log\left( \sqrt{\tilde{r}} + \sqrt{\tilde{r}-1}\right) - \sqrt{1- \frac{1}{\tilde{r}}} } {\left( 1 - \frac{1}{\tilde{r}} \right)^{3/2}}\textrm{,} \\
    \omega^{(C)} &=  \frac{3}{4} \frac{(\tilde{r}+2 \tilde{r} \log (\tilde{r})) D\left(\sqrt{\log (\tilde{r})}\right)-\tilde{r} \sqrt{\log (\tilde{r})}}{\left[\log(\tilde{r}) \right]^{3/2}}  \textrm{,}
\end{split}
\end{align}
for the spherical and cylindrical diodes correspondingly, where $\tilde{r}$ is the ratio between the emitter and the collector radii and $D(\cdot)$ denotes the Dawson function \cite{dawsonfun}.
The derivation is given in appendix \ref{sec:derivation}.

It is worth noting that for $\tilde{r} \rightarrow 1$, i.e. when the emitter and the collector have similar sizes, $\omega^{(S)} \approx 1 + \frac{2}{5}(\tilde{r}-1) $, and $\omega^{(C)}\approx  1 + \frac{1}{5}(\tilde{r}-1)$.
This is expected due to the fact that the spherical and cylindrical diodes become similar to the planar one, when the electrodes have similar radii and the gap distance diminishes. 
Moreover, for $\tilde{r} \rightarrow \infty$, i.e. when the collector electrode becomes infinitely larger than the emitter, $\omega$ scales as $\log(4\tilde{r})$ for the spherical case and as $3\tilde{r} (\log\tilde{r})^{-2}$ for the cylindrical one.
This means that the cylindrical diode deviates from the planar one much faster for increasing $\tilde{r}$.

\begin{figure}[h!]
    \centering
    \includegraphics[width = \linewidth]{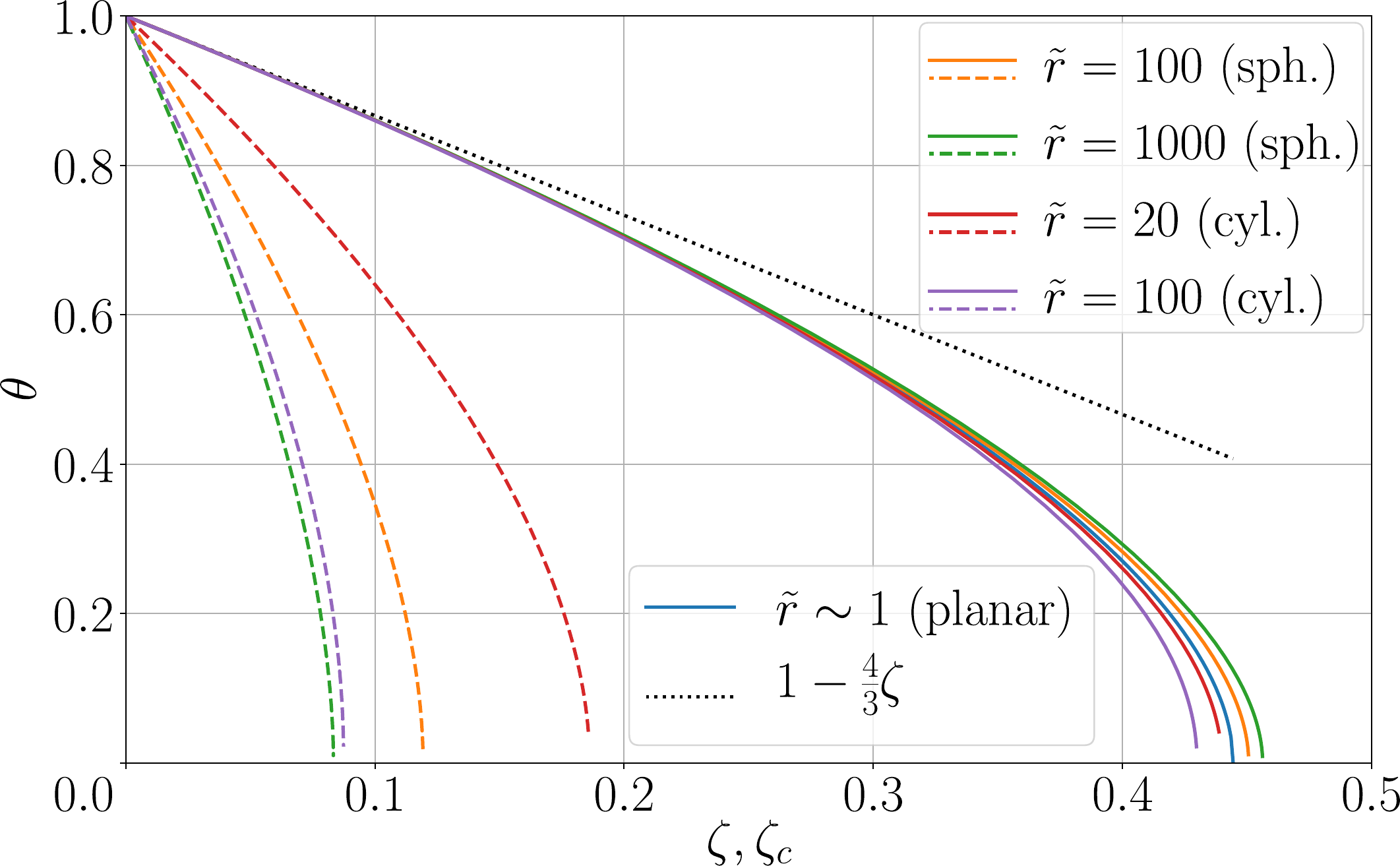}
    \caption{Field reduction factor $\theta=F/F_L$ vs $\zeta = kJ_s \sqrt{\chi}/F_L^{3/2}$ (dashed lines) and $\zeta_c = \omega \zeta$ (solid lines), for spherical and cylindrical field emitting diodes with various electrode radii ratios $\tilde{r} = r/R$. The linear approximation $1-\frac{4}{3}\zeta_c$ is shown for comparison.}
    \label{fig:zeta_theta_sph}
\end{figure}
In order to validate the above results, we solved the Poisson equation numerically by the Runge-Kutta method \cite{press1996numerical}.
The results are shown in figure \ref{fig:zeta_theta_sph} where we plot the field reduction factor $\theta$ vs both the SC strength $\zeta$ (dashed lines) and the corrected one $\zeta_c = \omega \zeta$ (solid lines), for various values of $\tilde{r}$, both for the cylindrical and the spherical diodes. 
It is evident from the dashed $\theta(\zeta)$ curves, that both the cylindrical and spherical geometries deviate significantly from the planar one and behave very differently depending on the radii ratio $\tilde{r}$.
However, after the correction $\zeta_c = \omega \zeta$, they all collapse in a curve that is very close to the one of the planar diode (blue).
Hence, the CEPD model can describe accurately these two curved geometries for a very wide range of parameters.
We note that the field suppression given by the CEPD model starts deviating significantly only for the strong SC regime, $\theta \lesssim 0.2$, while in the derivation of the model we aimed to fit only the linear term of eq. \eqref{eq:omega} (dashed line). Such an excellent agreement demonstrates the validity of the model far beyond the weak SC regime, which is limited to $\theta \gtrsim 0.6$.
Finally, note that the deviation increases with $\tilde{r}$, with the increase being much faster for the cylindrical diode.

\section{Methods} \label{sec:method}
\subsection{General computational method}
\label{sec:computational}

In this section we shall generalize the calculation of $\omega$, using a numerical method applicable to any geometry.
Our technique is based on the finite element code FEMOCS \cite{Veske2018}, which has been recently enhanced with PIC capabilities \cite{veske2019dynamic}.
In this work we do not aim to reproduce the temporal evolution of the SC distribution, but only the steady state, which is reached in a sub-picosecond timescale \cite{veske2019dynamic}.
In the steady state, the electric field distribution is constant and all particles emitted from a given point follow the same path to the collector.
Therefore, there is no need to solve the Poisson equation concurrently with the particle movement, but rather obtain a self-consistent solution of the electric field distribution and the particle trajectories.
A similar method has been used before by Zhu and Ang \cite{zhu2015space}.

Our method starts by considering no SC and solving the Laplace equation in the vacuum region, using the Finite Element Method (FEM), with a Dirichlet boundary condition $\Phi = 0$ at the emitter and $\Phi = V$ at the collector (see figure \ref{fig:schematic}).
Then the electric field distribution is calculated on the emitter surface and the field emitted current density is obtained utilizing the electron emission computational tool GETELEC \cite{GETELECpaper}, which evaluates the appropriate emission formulas.
Then charged superparticles (SPs) are injected from the surface into the vacuum.
Unlike our previous work \cite{veske2019dynamic}, here we inject one SP at the center of each surface element and adjust the corresponding SP weight according to the locally emitted current.
Thus the weight of the emitted SP is $w_{sp} = J_f A_f \Delta t/e$, where $J_f$ is the emission current density on a certain face element, $A_f$ is its area, $\Delta t$ is the path integration timestep, and $e$ the elementary charge. 

After injection, the electron paths are followed by numerically integrating Newton's equations of motion, until all electron SPs reach the anode boundary, where they are removed.
As in Ref. \cite{veske2019dynamic}, we use the explicit leapfrog method integration scheme \cite{dawson1983particle}.
However, since here we are integrating only the electron paths without a concurrent field and emission calculation, we can utilize an adaptive integration timestep. 
The latter starts at 0.1 fs upon particle injection, and is increased or decreased by 15\% (value found to ensure numerical stability) at some of the timesteps, in order to maintain an average of 2 timesteps that the particles stay in the same FEM cell.
This adaptive timestep technique ensures that near the emission surface, where the paths have high curvature and the charge density is high, the timestep is sufficiently low to give good accuracy. 
On the other hand, far from the emitter where the densities are very low and the paths have large radii of curvature, the integration remains at feasible CPU times, even for diode geometries where the electrodes have vastly different length scales (note that in figure \ref{fig:schematic}, $R / r_0 \simeq 2\times10^5$).
The decrease in the necessary CPU time is several orders of magnitude for such diodes.

At each integration timestep, the SPs give a contribution to the charge density (see eq. (5) and (6) of Ref. \cite{veske2019dynamic} for details), thus building up the SC distribution which is inserted in the assembly of the right hand side of the finite element Poisson equation.
The latter is then solved obtaining a new electric field distribution.
This process is repeated as a fixed point iteration, until convergence is reached, after typically 10-20 iterations.
After convergence, the electrostatic potential, field, charge density and current density distributions have been obtained.
As a convergence criterion, we demanded that the relative root mean square change on the charge density distribution is less than $10^{-4}$. 
We note that for the calculation of an $I-V$ curve, the voltage is gradually increased with small steps, utilizing the converged charge density of the previous step as the initial guess of the next.
This ensures the stable and fast convergence of the fixed point iteration method, as the initial guess is always relatively close to the desired convergence.

Finally, in order to obtain the CEPD parameter $\omega$ for a given point on the emitter surface, we calculate the distribution of $J_s, F,$ and $F_L$ on the emitting surface for various applied voltages $V$, using the numerical method described above.
For each point we can then calculate the corresponding $J_s - F_L$ curves of the CEPD model, by solving eq. \eqref{eq:Goncalves} self-consistently with the emission laws, using $d = \chi \omega^2$.
Then we optimize the value of $\omega$ in order to minimize the error between the CEPD and the numerical curves.

\subsection{Geometrical model for comparison to experiment}
\label{sec:socmodel}

Although our theory and computational methods are general with respect to the kind of emitted particles (electrons or ions) and the electrode geometry, we shall now focus on a specific example problem of field electron emission, for which experimental data are available for comparison.
In the experiments of Ref. \cite{BarbourSC}, an electrochemically etched tungsten cathode was utilized to take $I-V$ measurements at both low and high currents. 
The cathode was then coated with Ba in various coverages, in order to reduce the work function $W$ and reach SC-relevant current densities at achievable applied voltages. 

The shape of the cathodes used by the Linfield group \cite{BarbourSC, Dyke1953Arc, dyke1956field, DolanII, Dyke1953I}, including the one of Ref. \cite{BarbourSC}, are generally described by the general sphere-on-a-cone (SOC) model, developed by the same group \cite{DykeHemiCone}. 

\begin{figure}[h!]
    \centering
    \includegraphics[width = \linewidth]{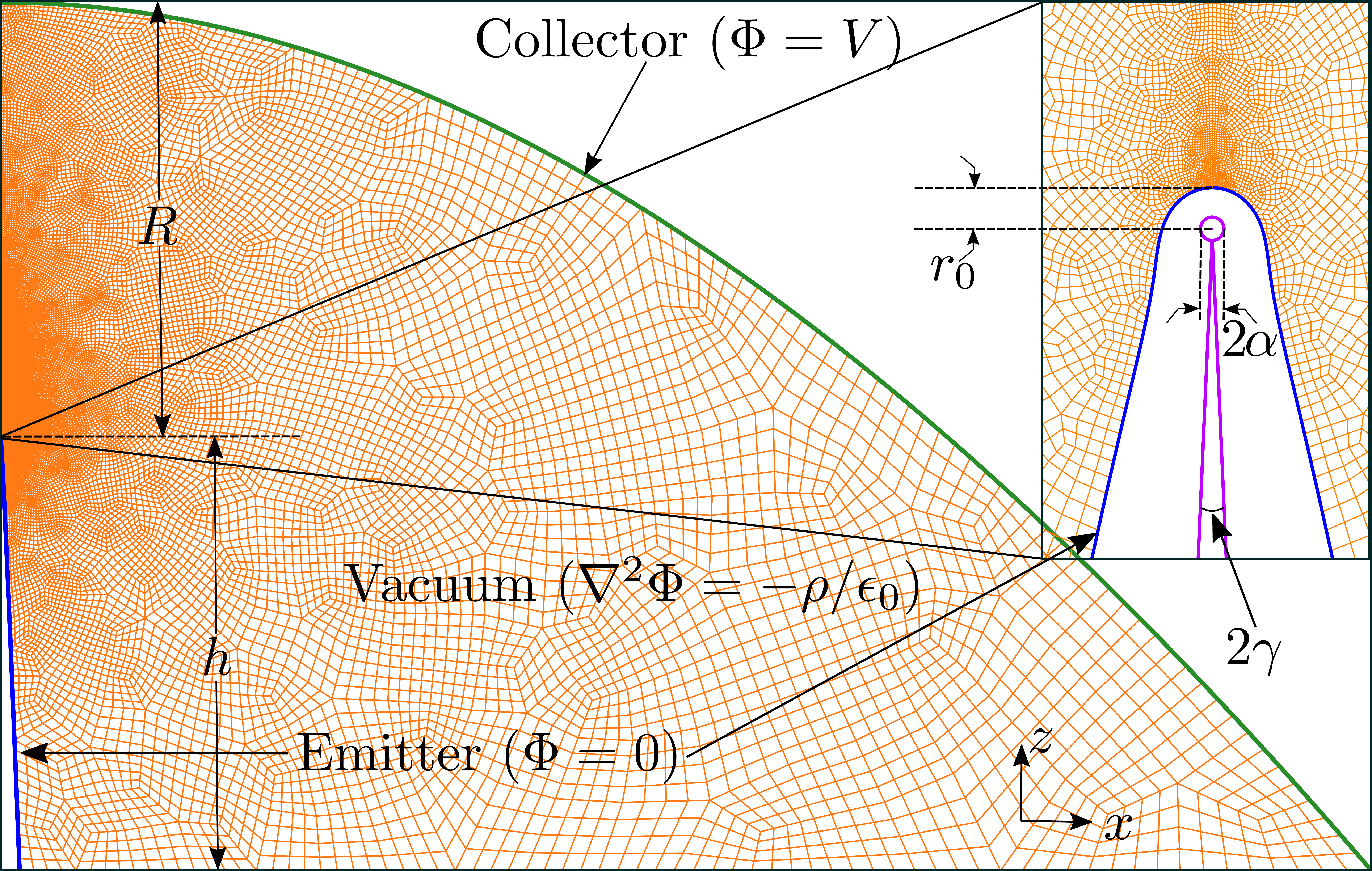}
    \caption{Schematic of the simulated system along with the quadrangular tesselation used in FEM.}
    \label{fig:schematic}
\end{figure}
In this model, the electrodes are shaped as equipotential surfaces of the electrostatic potential produced by an isolated charged sphere-on-a-cone electrode.
Such surfaces are defined by
\begin{equation} \label{eq:SOCpar}
    \left(r^n - \alpha^{2n+1}r^{-n-1} \right)P_n(\cos(\theta_p)) = C \textrm{,}
\end{equation}
where $(r,\theta_p)$ are the spherical coordinates (radius and polar angle correspondingly), $P_n(\cdot)$ denotes the Legendre function of the first kind \cite{legendreFun} of order $n \in (0,1)$, $\alpha$ is the radius of the sphere on the cone and $C$ is a constant parameter that determines which equipotential surface is defined by the equation.
The value of $n$ is determined by the aperture angle of the cone $\gamma$, via the relationship 
\begin{equation} \label{eq:Pleg}
    P_n(\cos(\pi - \gamma)) = 0 \textrm{.}
\end{equation}
The shapes of the two electrodes are determined by choosing two values of the parameter $C$.
The latter is determined from the desired radii of curvature at the apex ($\theta_p = 0$) of each electrode, $r = r_0$ for the emitter and $r = R$ for the collector, by evaluating eq. \eqref{eq:SOCpar}.

Figure \ref{fig:schematic} gives a comprehensive schematic of the model, along with the quadrangular tesselation we used to solve the Poisson equation by FEM.
The blue line corresponds to the emitter (cathode) surface, while the green one to the collector (anode).
In the inset at the upper right corner we zoom (several orders of magnitude) in the emitter apex region, where the virtual SOC electrode that defines the model geometry is shown in a magenta line. 

Barbour \textit{et. al.} \cite{BarbourSC} took micrographs of their tip and fitted a SOC model to it, however they did not explicitly report the parameters they extracted.
For this reason, to simulate their tip, we used the standard parameters described in Ref. \cite{DykeHemiCone} $n = 0.1$ (corresponds to $\gamma = 0.78^o$), $\alpha / r_0 = 0.235$, $R = 6.5$ cm, and $h = R$, apart from the scale parameter $r_0$, which was fitted to the experimental $I-V$ data, resulting in a value $r_0 = 315$ nm.
Although a direct comparison of these parameters to the emitter shape extracted by Barbour \textit{et. al.} from their micrograph is not possible, an indirect comparison gives very good agreement.
Our simulated geometry produces a theoretical field conversion factor (eq. (3) in Ref. \cite{DykeHemiCone}) $\beta \equiv \chi^{-1} =  0.3617 \textrm{ } \mu\textrm{m}^{-1}$, which is very close to the value $0.37 \pm 0.06 \textrm{ } \mu\textrm{m}^{-1}$ reported by Barbour \textit{et. al.} for the geometry they extracted from micrographs.
Note that the theoretical value of $\beta$ is slightly lower than the one calculated numerically, because the theoretical electrode SOC shapes extend to infinity.
Nevertheless, to approximate the real electrode shapes, we used a finite height $h = R$, which produces a slightly higher field enhancement.
The value of $h$ was chosen in view of the realistic electrode setup, as depicted in figure 2 of Ref. \cite{BarbourSC}.

\section{Results} \label{sec:results}

\begin{figure}[!ht]
    \centering
    \includegraphics[width = \linewidth]{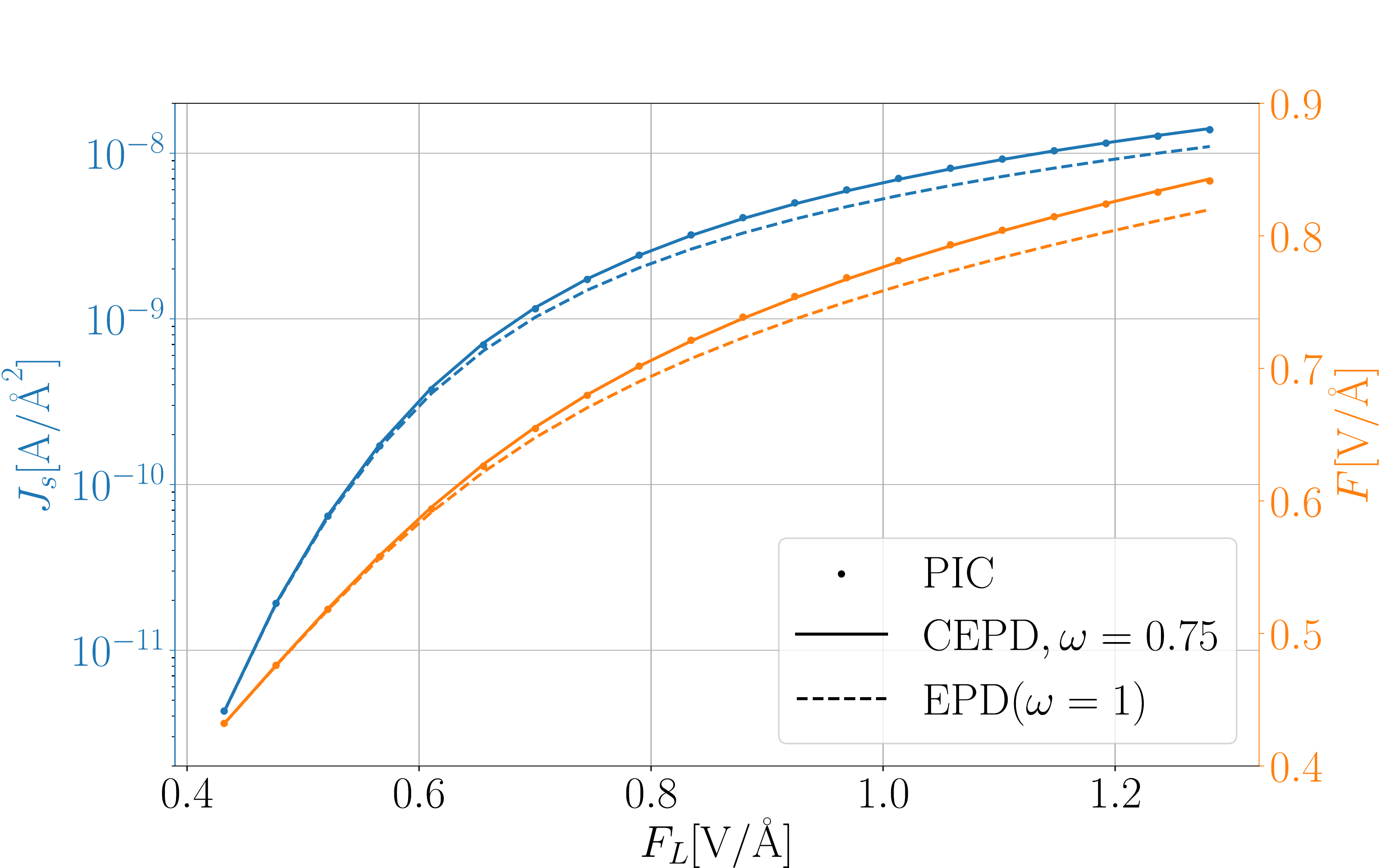}
    \caption{Emitted current density $J_s$ (left axis, blue) and local electric field $F$, as a function of the corresponding local Laplace field $F_L$, for a point on the emitting surface with direction $\theta_p = 45^o$, as calculated numerically (markers), according to the CEPD method with $\omega = 0.75$ (solid lines) and the simple EPD method ($\omega = 1$, dashed lines).}
    \label{fig:JF}
\end{figure}

Given the geometrical model described in section \ref{sec:socmodel}, we used the computational method described in section \ref{sec:computational} to solve the space charge problem and calculate the emission current density, charge density, and potential distributions, for various values of the applied voltage $V$.
In figure \ref{fig:JF} we plot the emitted current density $J_s$ and the surface electric field $F$ as a function of the Laplace field $F_L$ at a point on the emitter surface with polar angle $\theta_p = 45^o$.
The numerical results are shown in dots, while the CEPD and EPD calculations are shown in solid and dashed lines correspondingly.
We see that the CEPD model follows the PIC results very accurately, while the EPD model deviates significantly.

The value of the CEPD correction factor $\omega$, which is calculated by fitting to the PIC results as described in section \ref{sec:computational}, varies on the emitting surface, mainly due to the variation of the current density distribution $\xi(\tilde{\mathbf{r}})$. 
However, this variation of $\omega$ is small in the emission area (less than 10\%), allowing to use a single effective value $\omega_{\textrm{eff}}$ that describes the whole emitter geometry, without significant loss of accuracy.
The value of $\omega_{\textrm{eff}}$ is fitted by minimizing the deviation between the total current -- voltage ($I-V$) curve as calculated numerically and as estimated by the CEPD model.
The total current $I$ is calculated by numerically integrating $J_s$ over the emission surface for both cases.

\begin{figure}[!ht]
    \centering
    \includegraphics[width = \linewidth]{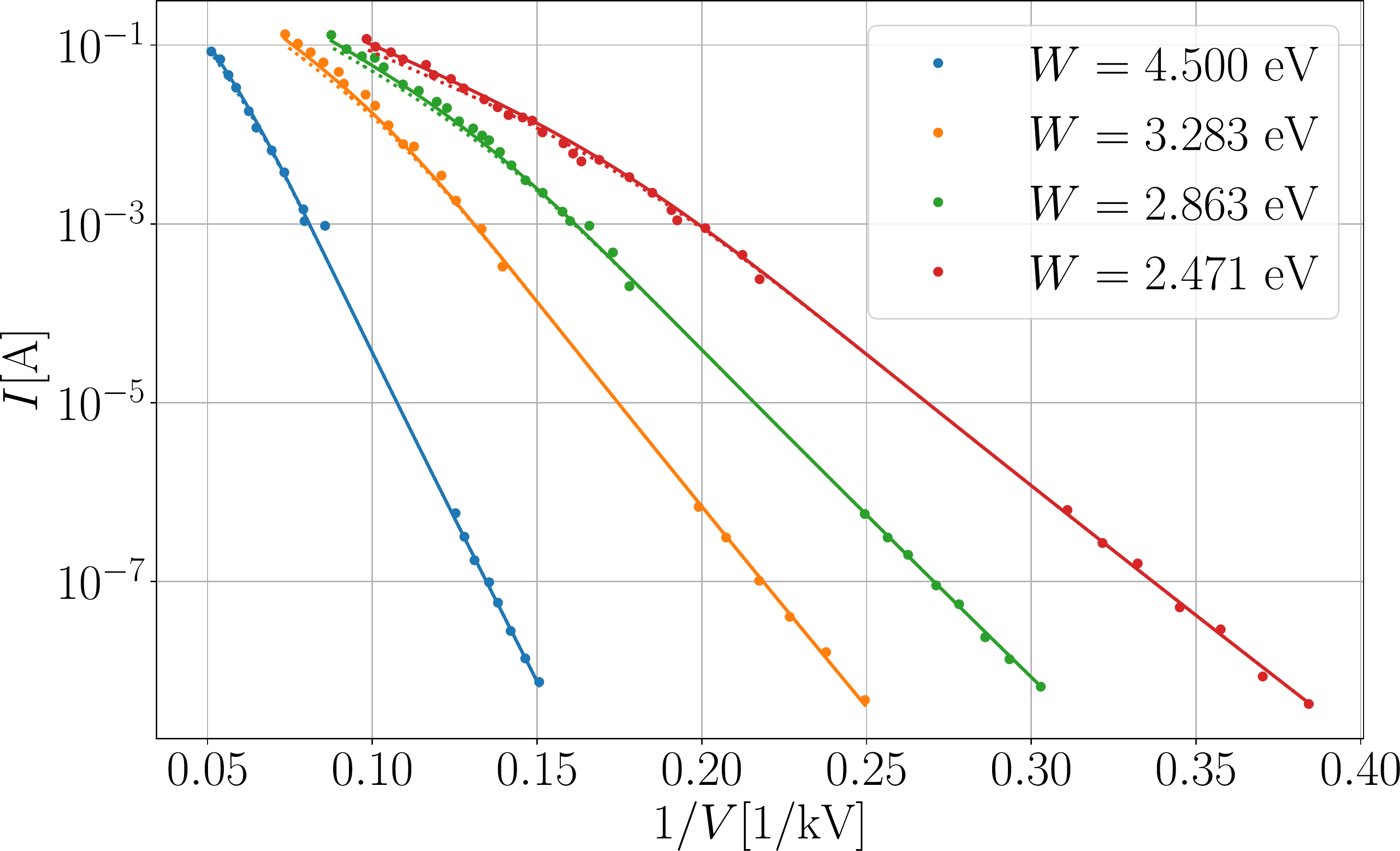}
    \caption{Experimental $I-V$ curves for various work functions for the emitter used in Ref. \cite{BarbourSC} (dots), along with the corresponding calculations using the CEPD model with $\omega_{\textrm{eff}} = 0.8$ (solid lines) and the standard EPD model (equivalent to $\omega_{\textrm{eff}} = 1$) are shown in dashed lines for comparison.}
    \label{fig:barbour_comp}
\end{figure}
In figure \ref{fig:barbour_comp}, we compare the experimental data of Ref. \cite{BarbourSC} (dots) with our calculations of the total current using the CEPD model with a single effective correction factor $\omega_{\textrm{eff}}$. 
We chose $r_0 = 315$ nm by fitting the theoretical curve for work function $W=4.5$ eV to the experimental data for a clean tungsten cathode, in the low-field regime where SC effects are negligible.
$r_0$ determines both the conversion length $\chi$ (equivalently the enhancement factor $\beta$), i.e. the slope of the curve, and the effective emission area, i.e. the vertical shift of the curve. 
We see that the value $r_0 = 315$ nm, which yields a calculated conversion length $\chi = 2.353 \textrm{ }\mu$m, produces an almost perfect match to the measurements.
Furthermore it is compatible with the shape extracted by the emitter micrographs in Ref. \cite{BarbourSC}.
Although the work function varies even on the surface of a clean emitter, we assumed a uniform effective work function $W$ for all the curves, indicated in the legend.
For the clean surface we assumed the standard tungsten value of $4.5$ eV, in line with Ref. \cite{BarbourSC}. 
The corresponding values for the coated cathodes, for which no prior knowledge is available, were fitted to match the measurements in the low field regime, after choosing $r_0$ from the clean-surface curve.
Finally, we note that for the lowest work function case, the current was multiplied by a fitted correction factor of 0.78, to account for the reduction of the effective emission area due to the increased non-uniformity of the work function, evident from the corresponding micrograph of figure 5 in Ref. \cite{BarbourSC}.
We then used the numerical method described above to calculate the CEPD model correction factor for this emitter geometry, yielding a value $\omega_{\textrm{eff}} = 0.8$.
Note that this value is slightly higher than the specific value of $\omega=0.75$ extracted for the point with $\theta_p = 45^o$ in figure \ref{fig:JF}, because $\omega_{\textrm{eff}}$ is an effective average of all the surface points.

We see that the CEPD model predicts very accurately the curvature of the experimental plots caused by the SC effects at high fields.
Surprisingly, the standard EPD model as introduced in \cite{BarbourSC}, gives also good agreement with the measurements.
This is due to the fact that $\omega_{\textrm{eff}}$ is very close to unity, resulting in a minor deviation between the EPD and CEPD models, as seen by the hardly distinguishable dashed curves. 

\begin{figure}[!ht]
    \centering
    \includegraphics[width = \linewidth]{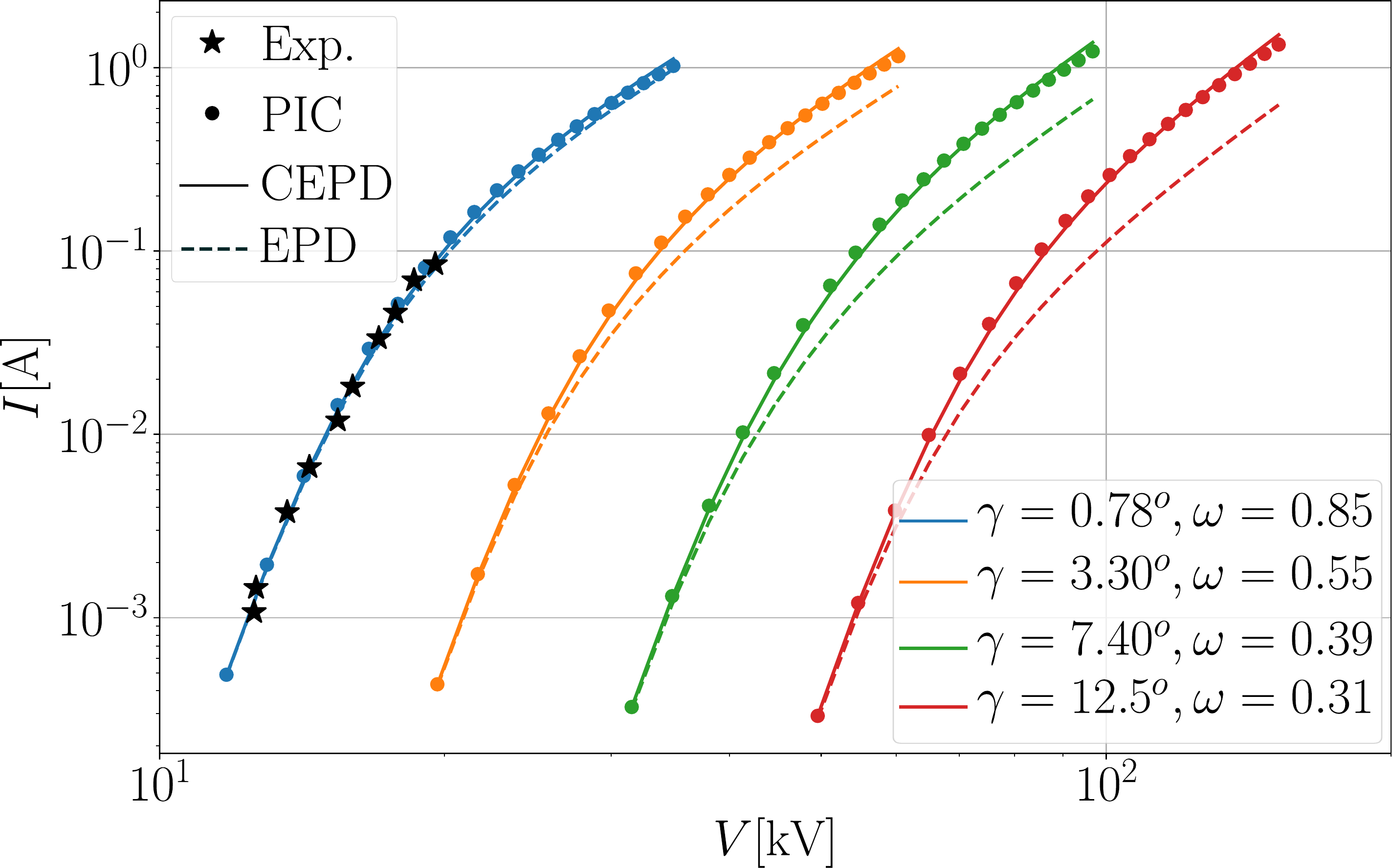}
    \caption{Comparison of $I - V$ curves for emitters of various cone apertures $\gamma$ as calculated by PIC (dots) and by the CEPD (solid lines) and EPD (dashed lines) models. The calculated correction factor $\omega_\textrm{eff}$ of the CEPD model is given in the legend for each geometry. The experimental data \cite{BarbourSC} for the clean emitter are given for comparison.}
    \label{fig:var_nleg}
\end{figure}
However, this is rather a coincidence, specific to this particular geometry.
In figure \ref{fig:var_nleg} we plot $I-V$ curves for different geometries, varying the cone aperture $\gamma$, while all other geometrical parameters are kept equal to the ones of figure \ref{fig:barbour_comp}. 
We see that as $\gamma$ increases, $\omega_\textrm{eff}$ decreases along with the field enhancement factor. 
This results in a significant deviation of the EPD model from the PIC calculations, while the CEPD model agrees with PIC much better.

Finally, in table \ref{tab:omegas} we tabulate the calculated values of $\omega_{\textrm{eff}}$ for various angles $\gamma$ and three different diode scales, i.e. different radii $r_0$, $\alpha$ and $R$, keeping constant ratii $R/r_0 = 10^4$ and $\alpha/r_0 = 0.235$.
The numerical results confirm the scale invariance of $\omega$ shown analytically by eq. \eqref{eq:omega}. We see that $\omega_{\textrm{eff}}$ has minute variations --within error margins-- for different diode scales.
The error margins are  obtained from the covariance matrix of the curve fitting to the PIC $I-V$ data.

\begin{table}
\begin{tabularx}{\linewidth}{l| X X X}
    \backslashbox{$\gamma$}{$r_0$} & 0.3$\mu$m &  1$\mu$m &  3$\mu$m \\
    \hline
    1$^o$ &0.951$\pm$0.020  & 0.945$\pm$0.021 & 0.935$\pm$0.021 \\
    2$^o$ & 0.754$\pm$0.016 & 0.748$\pm$0.017 & 0.739$\pm$0.021 \\
    4$^o$ & 0.564$\pm$0.014 & 0.560$\pm$0.015 & 0.552$\pm$0.015 \\
    8$^o$ & 0.478$\pm$0.012 & 0.473$\pm$0.013 & 0.466$\pm$0.013 \\
    16$^o$ & 0.348$\pm$0.010 & 0.335$\pm$0.010 & 0.329$\pm$0.010 \\
\end{tabularx}
\caption{$\omega_{\textrm{eff}}$ calculated for various geometries. Different values of $R$ correspond to different scales, i.e. keeping $\alpha/R=0.235$ and $R/r_0 = 10^4$.}
    \label{tab:omegas}
\end{table}

\color{black}
\section{Discussion}
The results of figures \ref{fig:barbour_comp} and \ref{fig:var_nleg} demonstrate the basic utility of the CEPD model. 
If the value of $\omega_\textrm{eff}$ and the field conversion length $\chi$ are available for a certain electrode geometry, the complex problem of calculating the SC suppressed emission current from a 3D emitter is reduced from running full PIC simulations, to evaluating the algebraic formula \eqref{eq:jseries} self-consistently with the emission characteristics (e.g. the Fowler-Nordheim equation).

In order to obtain $\omega_\textrm{eff}$ theoretically, PIC simulations are required, from which the CEPD model is fitted to the numerical $I-V$ curve.
However, this needs to be done only once for a given geometry; then the CEPD model can be used to calculate the emission current for different emission characteristics, such as work function and temperature.
This significantly reduces the computational cost of introducing emission routines that include the space-charge effects into more complex simulation techniques (see. e.g. \cite{kyritsakis2018thermal}). 

Furthermore, $\omega_\textrm{eff}$ values can be calculated and tabulated for typical emitter geometries becoming readily available for use.
The scale invariance of $\omega_\textrm{eff}$, demonstrated analytically in section \ref{sec:general} and numerically in table \ref{tab:omegas}, facilitates this by reducing the free parameters of any geometry.
Such a tabulation is out of the scope of this work and shall be given in a forthcoming publication.
Alternatively, exactly the same fitting procedure can be performed directly to the experimental $I-V$ curve instead of a numerical one, extracting directly $\omega_\text{eff}$ and giving a direct comparison between experiment and theory.

Finally, a comment is worthy regarding the generality of utilizing a single effective value $\omega_\textrm{eff}$ for the whole emitter.
As mentioned earlier, $\omega$ is defined separately for each point on the emitting surface, but if its variance within the area where the bulk of the emission originates is small, the usage of a single value $\omega_\textrm{eff}$ is sufficient. 
The same argument holds for the conversion length $\chi$ (or the field enhancement factor), allowing for the usage of a single $\chi$ value for an emitter, which is typical in the field emission community.
This approximation does not hold in case of emitters that have more than one distinct regions contributing significantly to the emission, with different geometrical characteristics each.
In this case, the approximation of a single effective value for $\omega$ is expected to become invalid, similarly to the invalidation of the standard Fowler-Nordheim theory for such emitters (see. e.g. Ref. \cite{popov2020influence}).

\color{black}
\section{Conclusions}
We have developed a three-dimensional theoretical model, describing the scaling laws of space charge limited charge emission at high electric fields.
Our model generalizes the one-dimensional planar model to be applicable for any geometry, using a geometry-specific correction factor. 
We validated our model by comparing it to both numerical calculations and existing experimental field emission data, either of which can be used to obtain the geometrical correction factor of our model.
We showed that the classical planar model tends to significantly overestimate the space charge effects, whereas our generalized theory is in very good agreement with both numerical calculations and experimental measurements.

\section*{Acknowledgments}
The current study was supported by CERN's CLIC K-contract No. 47207461 and the European Union’s Horizon 2020 program, under grant No 856705 (ERA Chair ``MATTER''). We also acknowledge grants of computer capacity from the Finnish Grid and
Cloud Infrastructure (persistent identifier urn:nbn:fi:research-infras-2016072533).

\appendix

\section{Derivation of the CEPD for the spherical and cylindrical geometries}

\label{sec:derivation}

In both the spherical and cylindrical geometries, the solution of the continuity equation can be obtained from the Gauss law, yielding $\xi(r)=(R/r)^n$, where $r$ is the radial coordinate, $R$ is the radius of the emitter, and $n=1,2$ for the cylindrical and spherical case correspondingly.
Then the Poisson equation becomes
\begin{equation} \label{eq:psphere}
    \frac{d}{dr}\left( r^n \frac{d\Phi}{dr} \right) = \frac{kJ_sR^n}{\sqrt{\Phi}} \textrm{,}
\end{equation}
where $J_s$ is the current density at the emitter surface, which is uniform due to symmetry.
For this problem, it is more convenient to solve the equivalent initial value problem (IVP) rather than the boundary value problem (BVP) addressed previously. 
Thus, we obtain the potential at radius $r$ as a function of the field on the emitter $F$.
Then this function is inverted in order to obtain $F$ as a function of a fixed potential $\Phi(r)$, which corresponds to an applied voltage $V = \Phi(r)$ at an electrode residing at $r$.

For this purpose, we use the reduced variables $\tilde{r} = r/R$, $\phi = \Phi/FR$ and solve the IVP
\begin{align} \label{eq:psphscaled}
\begin{split}
    \frac{d}{d\tilde{r}}\left( \tilde{r}^n \frac{d\phi}{d\tilde{r}} \right) & = \frac{kJ_s\sqrt{R}}{F^{3/2}} \frac{1}{\sqrt{\phi}} = \frac{\lambda}{\sqrt{\phi}} \textrm{,}    \\
    \left. \frac{d \phi}{d\tilde{r}} \right\rvert_{\tilde{r}=1} & = 1 \textrm{.}
\end{split}
\end{align}
The parameter $\lambda \equiv kJ_s\sqrt{R}F^{-3/2}$, as well as $\zeta$, is indicative of the SC strength.
We shall call $\lambda$ "implicit SC strength", in contrast to the explicit one $\zeta$, because $\zeta$ depends on the applied voltage $V$ (or equivalently on the Laplace field $F_L = V/\chi$). The latter is the a priori known free variable, unlike $F$, which is to be obtained.

The solution of \eqref{eq:psphscaled} can be expressed in terms of a Volterra integral equation of the second kind as
\begin{equation} \label{eq:volterra}
    \phi(\tilde{r}) = \phi_0(\tilde{r}) + \lambda \int_1^{\tilde{r}} \frac{\kappa(\tilde{r},\tilde{r}')}{\sqrt{\phi(\tilde{r}')}}d\tilde{r}' \textrm{,}
\end{equation}
where $\phi_0(\tilde{r})$ is the solution of the Laplace equation ($\lambda=0$), and the integration kernel $\kappa(\tilde{r}, \tilde{r}')$ is the Green function for the differential operator in the left hand side of eq. \eqref{eq:psphscaled}. 
It is
\begin{equation}
    \phi_0 = 1 - 1/\tilde{r}, \kappa = (\tilde{r} - \tilde{r}')/ \tilde{r}\tilde{r}'
\end{equation}
for the spherical case and 
\begin{equation}
    \phi_0 = \log(\tilde{r}), \kappa = \log(\tilde{r}/\tilde{r}')
\end{equation}
for the cylindrical one.
Eq. \eqref{eq:volterra} can be solved in the form of an asymptotic power series
\begin{equation}
    \phi = \phi_0 + \phi_1 \lambda + \phi_2 \lambda^2 + \cdots
\end{equation}
using the Adomian decomposition method.
The first-order term is obtained by inserting $\phi_0$ in the integral and yields
\begin{align}       \label{eq:phi1}
\begin{split}
    \phi_1^{(S)} & = \left(2- \frac{1}{\tilde{r}} \right) \log\left( \sqrt{\tilde{r}} + \sqrt{\tilde{r}-1}\right) - \sqrt{1- \frac{1}{\tilde{r}}} \textrm{,} \\
    \phi_1^{(C)} &= (\tilde{r}+2 r \log (\tilde{r})) D\left(\sqrt{\log (\tilde{r})}\right)-\tilde{r} \sqrt{\log (\tilde{r})}  \textrm{,}
\end{split}
\end{align}
for the spherical and cylindrical cases correspondingly.
In eq. \eqref{eq:phi1}, $D(\cdot)$ denotes the Dawson function \cite{dawsonfun}.

Now we shall use this to obtain an approximation for the Laplace field $F_L$ and the field reduction factor $\theta = F/F_L$. 
Given the potential $\Phi$ at distance $r$, the Laplace field on the emitter is $F_L = F \phi(\tilde{r}) / \phi_0(\tilde{r})$.
Hence, we can write
\begin{equation}
   F_L= F \left(1 + \lambda \frac{\phi_1}{\phi_0} + O(\lambda^2) \right)\textrm{,}
\end{equation}
which gives $F_L$ as a function of $(F,J,R,r)$ in the form of a Maclaurin series on $J$.
The inverse function $F(F_L,J,R,r)$ can be written in a similar series. 
Its first order term can be found by evaluating $\partial F/\partial J$, utilizing the implicit function theorem \cite{corwin1982multivariable}.
It yields 
\begin{equation}
    \label{eq:FofFL}
    F = F_L \left( 1 - \zeta \frac{\phi_1}{\phi_0^{3/2}} + O(\zeta^2) \right) \textrm{,}
\end{equation}
where $\zeta \equiv kJ_0\sqrt{R\phi_0}F_L^{-3/2}$, as defined in sec. \ref{sec:general}. By matching the coefficient of $\zeta$ in \eqref{eq:FofFL} and \eqref{eq:jseries}, we can obtain the correction factor of the CEPD for the spherical and cylindrical diodes as $\omega(\tilde{r}) = \frac{3}{4}\phi_1(\tilde{r})\phi_0^{-3/2}(\tilde{r})$, yielding eq.~\eqref{eq:omega_sph_cyl}.

\bibliography{bibliography/bibliography}

 \end{document}